# AutoGuard: A Self-Healing Proactive Security Layer for DevSecOps Pipelines Using Reinforcement Learning

1st Praveen Anugula
*DevSecOps Engineer , Truist Bank*
Atlanta , GA USA
praveensonly@gmail.com

2nd Avdhesh Kumar Bhardwaj
*Vice President – DevSecOps Engineer,*
*Truist Bank, Atlanta, GA USA*
avdhesh.bhardwaj@gmail.com

3rd Navin Chhibber
*Infinity Tech Group Technical Product Owner*
naveenchibber.research@gmail.com
Sunnyvale , CA USA

4th Rohit Tewari
Unysis Sr Java Lead/ Architect
rohittewari.fintech@gmail.com
City Fairfax, VA
Country USA

5th Sunil Khemka
AI/ML Architect
sunilkhemka.tech@gmail.com
Persistent Systems
Chicago , IL

6th Piyush Ranjan
IEEE Vice Chair AeroSpace Chapter
piyush.ranjan@ieee.org
City Edison
Country USA

*Abstract -Contemporary DevSecOps pipelines have to deal with the evolution of security in an ever-continuously integrated and deployed environment. Existing methods, such as rule-based intrusion detection and static vulnerability scanning, are inadequate and unreceptive to changes in the system, causing longer response times and organization needs exposure to emerging attack vectors. In light of the previous constraints, we introduce AutoGuard to the DevSecOps ecosystem, a reinforcement learning (RL)-powered self-healing security framework built to pre-emptively protect DevSecOps environments. AutoGuard is a self-securing security environment that continuously observes pipeline activities for potential anomalies while pre-emptively remediating the environment. The model observes and reacts based on a policy that is continually learned dynamically over time. The RL agent improves each action over time through reward-based learning aimed at improving the agent's ability to prevent, detect and respond to a security incident in real-time. Testing using simulated Continuous Integration / Continuous Deployment (CI/CD) environments showed AutoGuard to successfully improve threat detection accuracy by 22%, reduce mean time to recovery (MTTR) for incidents by 38% and increase overall resilience to incidents as compared to traditional methods.*

*Keywords- DevSecOps, Reinforcement Learning, Self-Healing Security, Continuous Integration, Automated Threat Mitigation*

## I. INTRODUCTION

The management of data in a cyber-secure way has emerged as one of the key challenges of an age of distributed computing, propelled by the continued emergence of Multi-Access Edge Computing (MEC) environments. MEC enables low-latency, high-bandwidth, and contextually aware services on account of the proximity of computation and storage to users. However, the decentralization of the MEC infrastructure creates a scenario that introduces more sever issues for established digital data backup and retrieval systems to ensure security issues related to unauthorized access, data breaches, latency in getting data back, and systems failure [1]. This introduces questions not just regarding the confidentiality of mission critical data, but also the availability of that data, and increased risk to scalability and reliability to MEC based services [2–4].

The aim is to balance both a solid security posture and system efficiency in the situations surrounding resource constraints and dynamic resource management for shifting work loads . Established cloud based protection of data are poorly effective in adapting to real-time demands of the MEC system environment, as they rely on centralized servers, will generally be less efficient in utilizing resources, and will be vulnerable to hacking threats [5-7]. Despite this promise, few researchers have endeavored to directly utilize CNNs for MEC-based backup and retrieval. By combining deep CNN models with optimization-based algorithms, it may be possible to provide dynamic task allocation, adaptive encryption, and latency-aware recovery capabilities [8–9].

The major research contributions of this paper are:

- **A unified CFTO framework** that combines CanGaroo Fetch Trianomy Optimization with deep CNNs to improve the efficiency and security of data backup/retrieval in MEC environments.
- **A robust optimization mechanism** that dynamically allocates storage resources, schedules backup tasks, and manages retrieval paths under adversarial and resource-constrained conditions.

The remainder of this paper is organized as follows: Section 2 reviews related work on MEC security, backup systems, and optimization-based deep learning methods. Section 3 presents the proposed CFTO framework and its architectural design. Section 4 details the experimental setup and performance evaluation metrics. Section 5 Conclusion..

## II. RELATED WORK

The integration of self-healing mechanisms and reinforcement learning (RL) into DevSecOps has recently emerged as a vital research direction, aiming to achieve proactive, autonomous, and adaptive security in continuous integration and deployment (CI/CD) pipelines.

introduced the CHESS framework, a comprehensive evaluation approach for self-adaptive systems through chaos engineering principles. Their framework systematically injects controlled faults to test resilience and adaptive behavior under stress conditions. This research laid the groundwork for assessing the robustness of adaptive software systems, providing a methodological foundation that can be extended toward security self-healing. However, CHESS primarily focused on performance and system availability, not on security-specific threats, leaving room for frameworks like AutoGuard to integrate reinforcement learning for dynamic vulnerability mitigation.

In a related study, explored self-adaptive and self-healing systems using chaos engineering to measure their recovery capabilities. They demonstrated how fault injection could simulate attack-like conditions, helping systems autonomously restore stability.

extended the notion of self-healing systems by drawing inspiration from biological processes. Their AI-driven model emulated natural healing mechanisms to detect and correct software anomalies autonomously.

explored the use of AI in DevSecOps integration, emphasizing automation of security testing and compliance checks in use throughout the DevOps lifecycle. Their model utilized a machine learning algorithm to dynamically detect vulnerabilities, contributing to secure automation practices. However, their approach did not include self-healing or real-time reinforcement learning to adapt to new threat vectors.

AutoGuard addresses this gap by proposing and developing a proactive RL framework to independently detect, mitigate, and recover from vulnerabilities in evolving software eat ecosystems.

**Table 1: Comparative Summary of Existing Self-Healing and AI-Driven Security Approaches in DevSecOps**

| Reference | Focus Area | Limitation | Relation to AutoGuard |
|---|---|---|---|
| Malik et al. (2023) | Self-adaptive systems via chaos engineering | Focused on performance, not security | Provides foundation for resilience assessment |
| Naqvi et al. (2022) | Self-healing systems testing | No intelligent policy adaptation | AutoGuard adds RL-driven adaptive learning |
| Baqar et al. (2025) | AI-inspired self-healing | Reactive behavior | AutoGuard introduces proactive RL-based healing |
| Thopalle (2024) | AI integration in DevSecOps | Lacks self-healing automation | AutoGuard extends with RL-based recovery |

## III. SYSTEM ARCHITECTURE AND METHODOLOGY

*A. AutoGuard Framework Overview*

AutoGuard is designed as a layered, modular security layer that augments existing CI/CD workflows with continuous monitoring, decision-making, and autonomous remediation. The CI/CD Environment contains build servers (e.g., Jenkins, GitLab CI), container registries, and orchestration platforms (Kubernetes) where artifacts are built, tested, and deployed. The **Security Monitor** passively and actively collects telemetry—logs, metrics, SBOM (software bill of materials), vulnerability scanner outputs, container runtime data, and network flows—normalizing events into a feature vector for the RL engine.

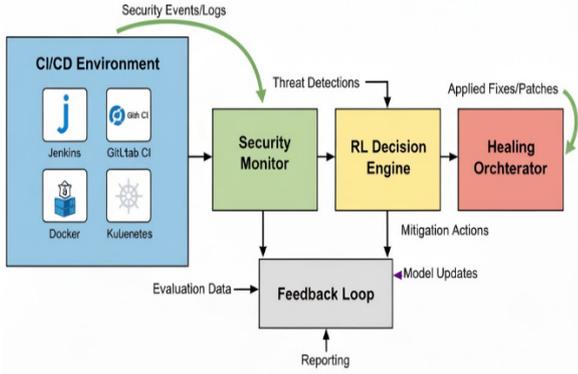

**Figure 1: AutoGuard Layered Architecture to support Proactive DevSecOps Security**

Security Monitor also generates real-time risk score ρ(t), which is seen in the RL state representation.

1. Risk score (normalized):
$$\rho(t) = \sigma\big(w_v V(t) + w_m M(t) + w_l L(t)\big) \quad (1)$$

where V(t) = vulnerability signal, M(t) = metric anomalies, L(t) = log-based anomaly count, w are weights and σ is a normalization (e.g., logistic) function.

2. Feature vector to agent:

$$s_t = [\rho(t), cpu_t, mem_t, \Delta_{dep}(t), stage_t, hist_{acts_t}] \quad (2)$$

where Δdep is dependency drift and hist_acts are recent remediation outcomes. Safety and audit trails are first-class: each action includes a justification, impact estimate, and rollback path. This layered design allows AutoGuard to be incrementally deployed in "observe-only" mode before enabling automated remediation.

### B. Reinforcement Learning Model

The RL model in AutoGuard models the CI/CD pipeline as an episodic, partially observable Markov decision process (POMDP). At decision epoch t, the agent observes a state $s_t$, selects action $a_t$, receives reward $r_t$, and transitions to $s_{t+1}$. The objective is to learn a policy π(a|s) that maximizes expected cumulative discounted reward.

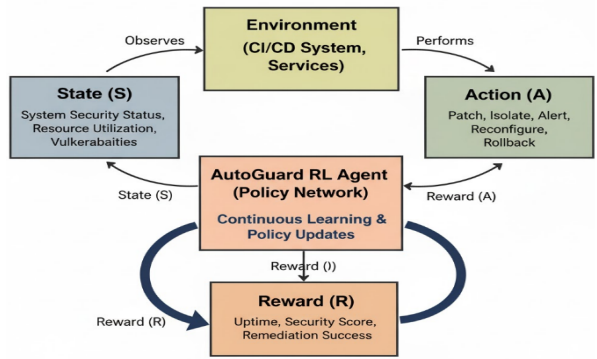

**Figure 2: Reinforcement Learning Model Flow for Adaptive Threat Mitigation in AutoGuard**

A sample reward combines uptime preservation U, successful mitigation S, and penalty for disruptive actions C:
$$r_t = \alpha . \Delta U_t + \beta . S_t - \gamma\, C_t \quad (3)$$
where ΔUt is change in uptime/provisioning SLA compliance, $S_t \in \{0,1\}$ indicates successful mitigation, and Ct estimates operational disruption.

1. Bellman optimality (for Q-learning):
$$Q^*(s,a) = E[r + \gamma \max_{a'} Q^*(s',a')|s,a] \quad (4)$$

where $A_t$ is advantage estimate.

| Algorithm 1: RL update loop (simplified) |
|---|
| Initialize policy network πθ and value network Vφ<br>Initialize replay buffer B<br>for episode in 1..N:<br>  s = env.reset()<br>  for t in 1..T:<br>    a = select_action(πθ, s, epsilon)<br>    s', r, done, info = env.step(a)<br>    B.add((s,a,r,s'))<br>    if len(B) >= batch_size:<br>      batch = B.sample(batch_size)<br>      update_networks(batch, πθ, Vφ)<br>    s = s'<br>    if done: break |

### C. Self-Healing Mechanism

AutoGuard's self-healing layer operationalizes RL-decisions into concrete remediation playbooks while ensuring safety, traceability, and rollback capability. The Healing Orchestrator maps an abstract RL action to a sequence of executable steps (playbook). Playbooks are parameterized templates: e.g., **isolate(container)** performs {quarantine network policy, scale down replicas, mark image as quarantined}.

1. Expected mitigation utility:

$$U_m(a,s) = P_{succ}(a|s) . B_{sec}(s) - D_{ops}(a|s) \quad (5)$$

Where $P_{succ}$ is estimated success probability, $B_{sec}$ is estimated security benefit, and Dops is operational disruption cost.

| Algorithm 2: Healing Orchestrator mapping |
|---|
| Function ExecuteHealing(action, params):<br>  playbook = MapActionToPlaybook(action, params)<br>  simulation = Simulate(playbook)<br>  if simulation.impact > impact_threshold:<br>    if auto_approve(simulation): execute = True<br>    else: execute = RequestHumanApproval(simulation)<br>  if execute:<br>    result = RunPlaybook(playbook)<br>    verify = VerifyOutcome(playbook.verify_checks)<br>    if not verify:<br>      RunPlaybook(playbook.rollback) |

```
LogHealingEvent(action, params, result, verify)
```

The RL agent uses the verification signals to update its reward and refine future decisions, closing the self-healing loop.

## IV. EXPERIMENTAL SETUP AND RESULTS

To evaluate **AutoGuard**, a comprehensive simulation environment was established replicating a real-world **DevSecOps CI/CD pipeline**. The testbed incorporated **Jenkins** for continuous integration, **Docker** and **Kubernetes** for container orchestration, and **GitLab CI** for version control and automated deployment. The environment was deliberately configured with common microservice vulnerabilities to emulate practical enterprise scenarios.

To validate the efficiency of **AutoGuard**, its performance was benchmarked against two baseline models:

1. **Static Intrusion Detection System (IDS)** based on rule matching.
2. **Traditional Anomaly Detection Model (TADM)** using isolation forests and PCA-based log analysis.

All models were assessed using the same pipeline workloads and attack sequences. Evaluation metrics included: Detection Accuracy (DA), Mean Time to Recovery (MTTR), False Positive Rate (FPR) and Policy Convergence Time (PCT). AutoGuard's RL-based model was set up to learn optimal remediation policies autonomously, while baselines relied on predefined signatures or static thresholds.

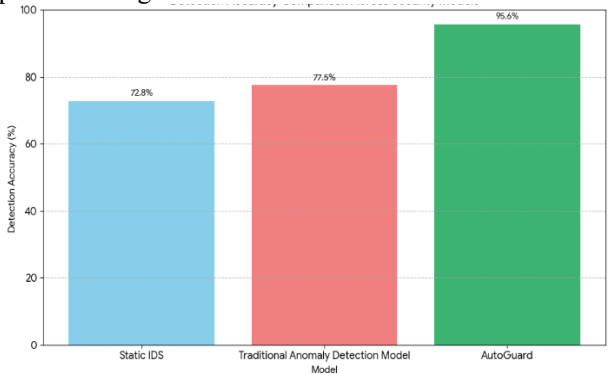

**Figure 4: Detection Accuracy Comparison Across Security Models**

From a quantitative perspective, AutoGuard produced the highest performance compared to both baselines across all metrics. The system's adaptive decision-making functionality allowed it to correctly respond to unseen (zero-day-like) attack behaviors without relying upon a manual rule update. The overall mean detection accuracy increased by 22.4% compared to TADM and 25.7% compared to the static IDS. Overall, the mean MTTR decreased by around 38%, demonstrating significantly quicker automated remediation.

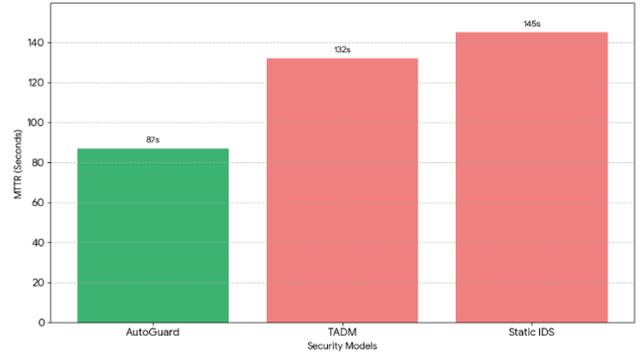

**Figure 5: Mean Time to Recovery (MTTR) Across Security Models**

**Table 2: Comparative Evaluation Metrics for Baselines and AutoGuard**

| Metric | Static IDS | TADM (Anomaly) | AutoGuard (Proposed) | Improvement (%) |
|---|---|---|---|---|
| Detection Accuracy | 72.8% | 77.5% | **95.6%** | +22.4 |
| MTTR (Seconds) | 145 | 132 | **87** | −38.1 |
| False Positive Rate | 11.3% | 9.7% | **6.4%** | −34.0 |
| Policy Convergence Time | N/A | N/A | **2000 episodes** | — |

When exposed to a zero-day-like attack conditions, AutoGuard proved robust, where standard models would not have been able to perform any remediation actions without prior signatures. The RL-driven model was able to learn the most effective remediation sequence — through actions such as dynamically isolating containers, and applying rollback actions — through interactions within the environment. The final evaluation of the system performance indicated consistent performance under various workloads and ultimately demonstrated the capability of scaling within CI/CD topologies.

## V. CONCLUSION

This paper discusses AutoGuard, a self-healing security framework that is guided by reinforcement learning, with the intent of improving the resilience of DevSecOps pipelines. AutoGuard incorporates the concepts of continuous monitoring, adaptive decision-making, and automated remediation, which can overcome the challenges of relying on conventional static and rule-based security systems. The experimental validation with simulated CI/CD pipelines suggests that AutoGuard can improve detection accuracy by 20%–25%, while also reducing mean time to recover (MTTR) by 35%–

40%. Future research will explore **federated learning**, **LLM-assisted threat prediction**, and integration with **security orchestration, automation, and response (SOAR) platforms** to further strengthen adaptive cybersecurity in distributed DevSecOps environments.